\documentclass[aps,prc,onecolumn,notitlepage,showpacs,12pt]{revtex4-1}
\usepackage{epsfig}

\usepackage{amsmath}
\usepackage{graphicx}
\usepackage{amssymb}
\usepackage{bm}
\long\def\symbolfootnote[#1]#2{\begingroup%
\def\thefootnote{\fnsymbol{footnote}}\footnote[#1]{#2}\endgroup}
\begin{document}

\title{Nuclear shell effect  and  collinear  tripartition of nuclei}
\author{A.K. Nasirov$^{1,2}$, W. \lowercase{von} Oertzen$^3$, R.B. Tashkhodjaev$^{2}$}

\affiliation{$^1$Joint Institute for Nuclear Research, Joliot-Curie 6, 141980 Dubna, Russia,\\
$^2$Institute of Nuclear Physics, Uzbek Academy of Science, 100214 Tashkent, Uzbekistan\\
$^3$Helmholtz-Zentrum Berlin, Glienickerstr. 100, 14109 Berlin, Germany\\
$^4$Fachbereich Physik, Freie Universit$\ddot{\rm a}$t, Berlin}

\begin{abstract}
A possibility of formation of the three reaction products having comparable masses at the spontaneous fission of $^{252}$Cf is theoretically explored. This work is aimed to study the mechanism leading to observation of the reaction products with masses $M_1=$136---140 and $M_2=$68---72 in coincidence by the FOBOS group in
JINR. The same type of ternary fission decay has been observed
 in the reaction $^{235}$U(n$_{\rm th}$,fff). The potential energy surface for the ternary system forming a collinear nuclear chain is calculated for the wide range of mass and charge numbers of constituent nuclei. The results  of the PES for the  tri-partition of $^{252}$Cf(sf,fff)  shows, that  we have  favorable dynamical conditions for the formation of fragments with  mass combinations of clusters
  $^{68-70}$Ni with  $^{130-132}$Sn and with missing cluster $^{48-52}$Ca.
\end{abstract}



\maketitle

% \tableofcontents
% \listoffigures
% \listoftables
\section{Introduction}
Binary fission has been studied intensively over the last four decades,
for an overview there are the books edited
 by R. Vandenbosch and J. R. Huizenga \cite{Vandenbosch},
  C. Wagemans \cite{wag91}, covering all
important aspects of this process. A more recent theoretical coverage is available as a textbook  by H. Krappe and K. Pomorski in ref.~\cite{Krappe}.
Ternary fission, when a third light particle is emitted perpendicular
to the binary fission axis, has also been studied extensively ~\cite{Gonnen04, Gonnen05}.
The name ``ternary'' fission has been used  so far for such decays by emission of light charged particles with mass numbers ($M<38$).
These ternary decays  give decreasing yields as
 function of increasing  mass(charge) of the third particle~\cite{Gonnen04}.
 The probability of the ternary fission by emission of the alpha particle relative to the binary fission is about $2\cdot 10^{-3}$ for the reactions ranging from  $^{229}$Th(n$_{\rm th}$,f) up to $^{251}$Cf(n$_{\rm th}$,f) \cite{Gonnen05}.

Recent experimental observations of the two fragment yields
in coincidences by the two FOBOS-detectors~\cite{Pyatkov10,Pyatkov12} placed at 180$^o$,
 using the  missing mass approach,
 have established the phenomenon of  \emph{collinear cluster tripartition} (CCT) of the massive nuclei.  This new decay mode has been observed
 for the spontaneous                                                                                                                                                                                                                                                                                                                                                                                                                                                                                  decay of  $^{252}$Cf(sf,fff) and for neutron induced
fission in $^{235}$U(n$_{\rm th}$,fff),  see refs.~\cite{Pyatkov10,Pyatkov12,WvOertzen13}.
In this CCT with the emission of three fragments, the outer fragments of the ``chain'' are registered ~\cite{Pyatkov10} only. The mass number of the missed third fragment can be larger than one of the heaviest light charged particle  of the ternary fission above mentioned. Therefore, the CCT process is called as
one of the mechanisms of the true ternary fission, when the masses of its products are relatively comparable.  The mass correlation plots $M_1-M_2$ ($M_1$ and $M_2$ are
mass numbers of products) of the registered reaction products showed the appreciable yield of magic isotopes of $^{68,70}$Ni, $^{80,82}$Ge, $^{94}$Kr,  $^{128,132}$Sn and $^{144}$Ba.
 These products were registered in the coincidence, but sum of their mass numbers
 differs from the total mass numbers $M_{\rm CN}$ of $^{252}$Cf and $^{236}$U:
$M_3=M_{\rm CN}-(M_1+M_2)$,  $4<M_3<52$.   $M_3$ is the mass number of the missed
fragment at registration.
It should be noted that the exotic fission products with mass numbers $61< M < 76$  (isotopes of Fe, Ni, Zn, Ge) have been observed as the very asymmetric fission products
\cite{Rao79,Barreau85,Sida89,Goverd95}.
Authors  of Ref. \cite{Goverd95} concluded that large deformation($\beta_2$=0.84) of the heavy fragment  $^{167}$Gd conjugate to $^{70}$Ni and transition through a potential barrier with
the wide width ($\Delta r$=4.5 fm) can explain the observed unusual small value of the kinetic energy of the light fission product of $^{236}$U(n,f) reaction ($E_{\rm n}$=1 MeV).
The aim of the present work is to analyze the formation of the $^{68,70}$Ni clusters in the true ternary fission of $^{252}$Cf and $^{235}$U(n$_{\rm th}$,f).

\section{True ternary fission}

The above mentioned experimental observations of the two fragment yields
in coincidences by the two FOBOS-detectors~\cite{Pyatkov10,Pyatkov12} placed at 180$^{\circ}$
have given an evidence of the true ternary fission, which was predicted  in the
theoretical works~\cite{DG,poe05,roy95, Manim} for long time ago.
The collinear configuration is preferred relative to the oblate
configuration for heavy system of ternary fragments with larger charges and masses
\cite{Manim}. In the last paper, the results of potential energy and relative yield
calculations reveal that
collinear configuration increases the probability of emission of heavy fragments
like $^{48}$Ca and its neighboring nuclei as the third fragment.
 The latter, Ca,  as the
smallest third particle is positioned  along  the line connecting Sn and Ni, in
this way minimizing the potential energy.
In the experiments described in Refs.~\cite{Pyatkov10,Pyatkov12},
two of the three fragments moving in the opposite directions were detected.
The  third fragment could not be registered due to the smallness of its velocity.
The range of the kinetic energy values was studied in Ref. \cite{Vijayaraghavan}.
  Thus  binary coincidences of the $^{68,70}$Ni isotopes and fission products
  with $A\approx 132$ are observed in the spontaneous fission of $^{252}$Cf and
  $^{235}$U(n$_{\rm th}$,f) reaction ~\cite{Pyatkov10,Pyatkov12}.
At the observation of $^{68,70}$Ni in the former and latter reactions
  the missed fragments are Ca and Si isotopes, respectively.
 To study theoretically the possibility of the formation of  $^{68,70}$Ni
 isotopes in coincidence with the  fission product having the mass number
  $A\approx$ 132---136, in the present work, the potential energy surface (PES) of the ternary system is calculated and analyzed.

\subsection{Potential energy surface for the ternary system}
 PES is found as a sum of the energy balance of the interacting fragments and nucleus-nucleus interaction
 between them
 \begin{eqnarray}
 \label{PES3}
&&U(R_{13},R_{23},Z_1,Z_3,A_1,A_3)=Q_{ggg}+V_{12}^{(Coul)} (Z_1,Z_2,R_{13}+R_{23})
\nonumber\\
&& +V_{13}(R_{13},Z_1,Z_3,A_1,A_3)+V_{23} (R_{23},Z_3,Z_2,A_3,A_2 ),
\end{eqnarray}
\begin{figure}
%\begin{center}
\vspace*{-0.75 cm}
\includegraphics[width=0.85\textwidth]{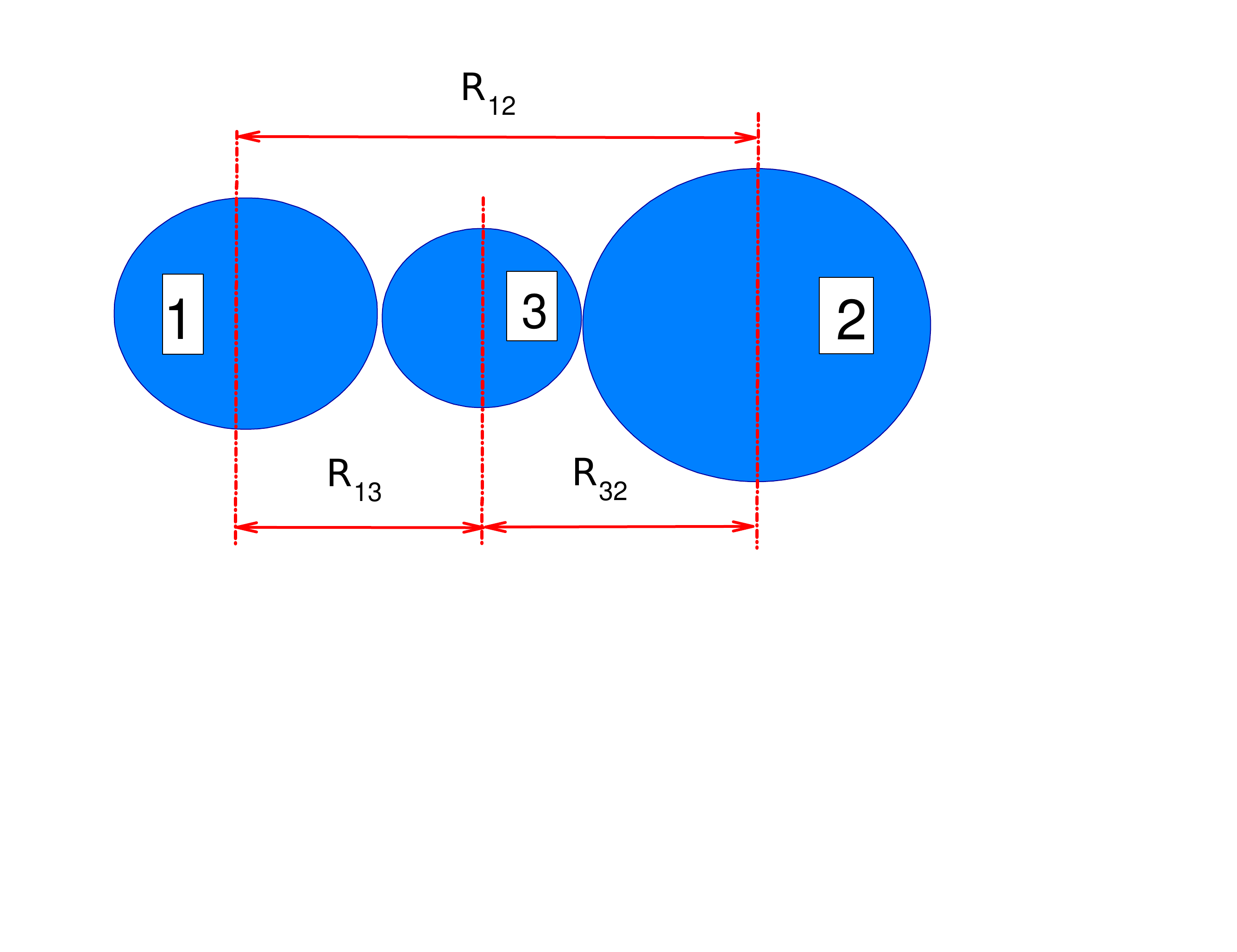}
\vspace*{-5.65 cm}
\caption{\label{ThreeFragRij} The relative distances between mass centres of a ternary system for collinear configuration.}
%\end{center}
\vspace*{-0.25 cm}
\end{figure}
\noindent
where   $Q_{\rm ggg}=B_1+B_2+B_3-B_{\rm CN}$ is the balance of the fragments binding energy at the ternary fission;  the values of binding energies are obtained
  from the mass table in Ref. \cite{Audi03}; $V_{13}$ and $V_{23}$ are
  the nucleus-nucleus interaction of the middle cluster ``3'' ($A_3$ and $Z_3$
  are its mass and charge numbers, respectively) with the left ``1''
 ($A_1$ and $Z_1$)  and right ``2'' ($A_2$ and $Z_2$) fragments of the ternary system; $V_{12}^{(\rm Coul)}$
 is the Coulomb interaction between two border fragments  ``1'' and ``2'',
  which are separated by the distance $R_{13}+R_{23}$, where $R_{13}$ and $R_{23}$ are the distances
  between the middle cluster and two outer clusters placed on the left and right sides, respectively (see Fig. \ref{ThreeFragRij}).
  The interaction potentials $V_{13}$ and $V_{23}$ consist of the Coulomb and nuclear parts:
 \begin{eqnarray}
   &&V_{3i}(R_{3i},Z_i,Z_3,A_i,A_3)=V_{3i}^{(\rm Coul)}(Z_i,Z_3,R_{i3})\nonumber\\
   &&+V_{3i}^{(\rm Nucl)}(Z_i,A_i,Z_3,A_3,R_{3i}), \hspace*{0.25cm} {\rm where} \hspace*{0.25cm} i=1,2.
 \end{eqnarray}
  The nuclear interaction calculated by the double folding procedure with the effective
  nucleon-nucleon forces depending on nucleon distribution density (see Ref. \cite{Tashk2011}).
  The Coulomb interaction is determined by the Wong formula \cite{Wong1973}.

Theoretical interpretation of the collinear tri-partition of  $^{252}$Cf and  $^{236}$U
\cite{Pyatkov10,Pyatkov12,WvOertzen13} needs the knowledge about the
mechanism of fission of the residual super-deformed system.
  The sequential mechanism of the true ternary fission without
 correlation between two ruptures of the two necks connecting three
 fragments in collinear configuration was assumed in Ref. \cite{Tashk2011}.
 The realization of the asymmetric fission channel as the first stage
   of sequential mechanism was considered. At the second stage the heavy product undergoes
   fission forming two nuclei with  comparable masses. The probability of
   fission depends on the fission barrier, which is very high for the relatively
   light nuclei. For example, the fission probability of the nuclei lighter
   than $^{158}$Ce formed with large probability at fission of actinides is very small
   (see Table 1 and Ref. \cite{Tashk2011}).
   To estimate the yield of the ternary fission products
   in different channels, we compared the yield of the reaction
   products in fission of   $^{144}$Ba, $^{150}$Ce and $^{154}$Nd
\begin{minipage}{16.5cm}
\small{{\bf Table 1}.  The realization probabilities of the different sequential
channels for the collinear cluster tripartition of $^{236}$U$^*$. ``*'' means
that these nuclei are excited.
%\vspace*{0.25cm}
\begin{center}
\begin{tabular}{|c|c|c|}
\hline
 Fission channel  & Fission channel of & Probability    \\
     $^{236}$U$^*\rightarrow f_1+f_2$ & primary heavy fragment&  of CCT      \\
\hline
$^{82}$Ge$^*$+$^{154}$Nd$^*$ &
$^{154}$Nd$^*\rightarrow^{72}$Ni$^*$+$^{82}$Ge$^*$& $3\cdot 10^{-4}$\\
                        & $^{154}$Nd$^*\rightarrow^{76}$Zn$^*$+$^{78}$Zn$^*$& $1.5\cdot 10^{-4}$\\
\hline
$^{86}$Se$^*$+$^{150}$Ce$^*$ &
$^{150}$Ce$^*\rightarrow^{66}$Fe$^*$+$^{82}$Ge$^*$& $ 1.0\cdot10^{-5}$\\
                        & $^{150}$Ce$^*\rightarrow^{72}$Ni$^*$+$^{76}$Zn$^*$& $1.4\cdot 10^{-5}$\\
\hline
\end{tabular}
\end{center}
}
\end{minipage}
\begin{figure}
\vspace*{-1.15 cm}
\resizebox{0.70\textwidth}{!}{\includegraphics{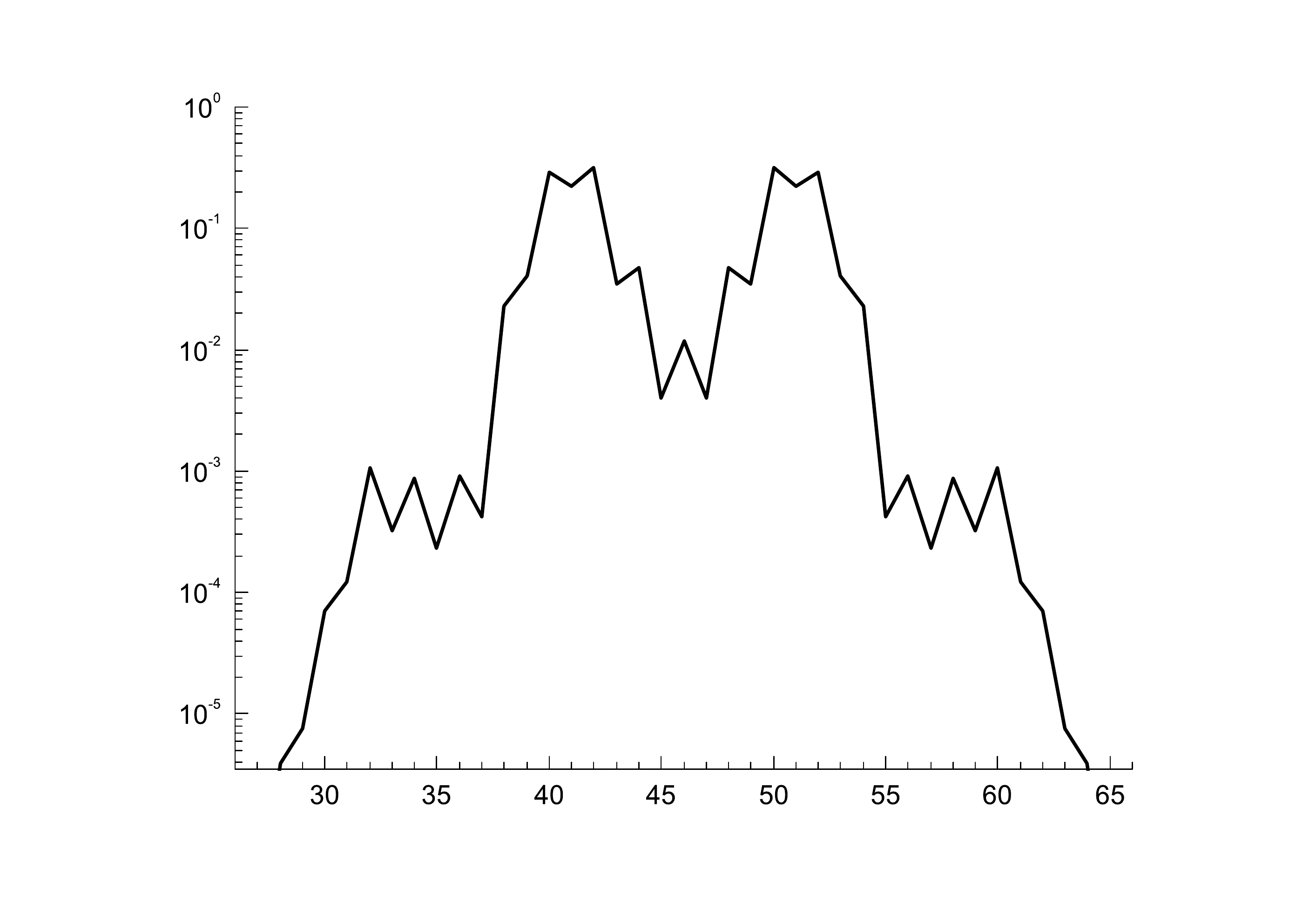}}
\vspace*{-1.15 cm}
\caption{\label{Y236U} Yields of the reaction products at the binary fission of $^{236}$U calculated
with the statistical method as in Ref. \cite{Tashk2011}.}
\end{figure}

\begin{figure}
\vspace*{-1.0cm}
\includegraphics[width=0.70\textwidth]{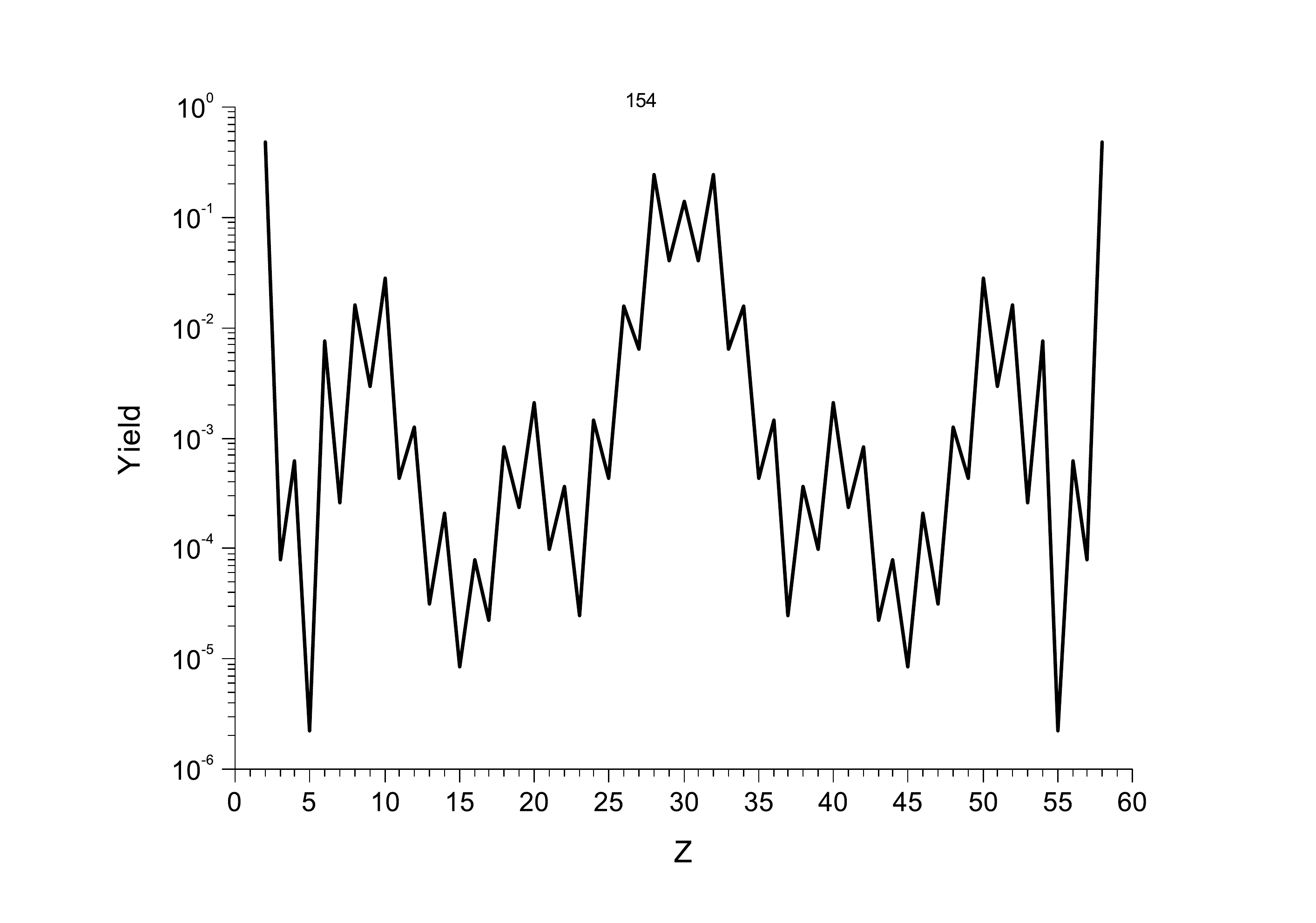}
\vspace*{-1.0cm}
\caption{\label{Y154Nd}Yields of the reaction products at the (sequential)
fission of $^{154}$Nd formed in the first step at binary fission of $^{236}$U.}
\vspace*{-0.25 cm}
\end{figure}

\begin{figure}
\includegraphics[width=0.80\columnwidth]{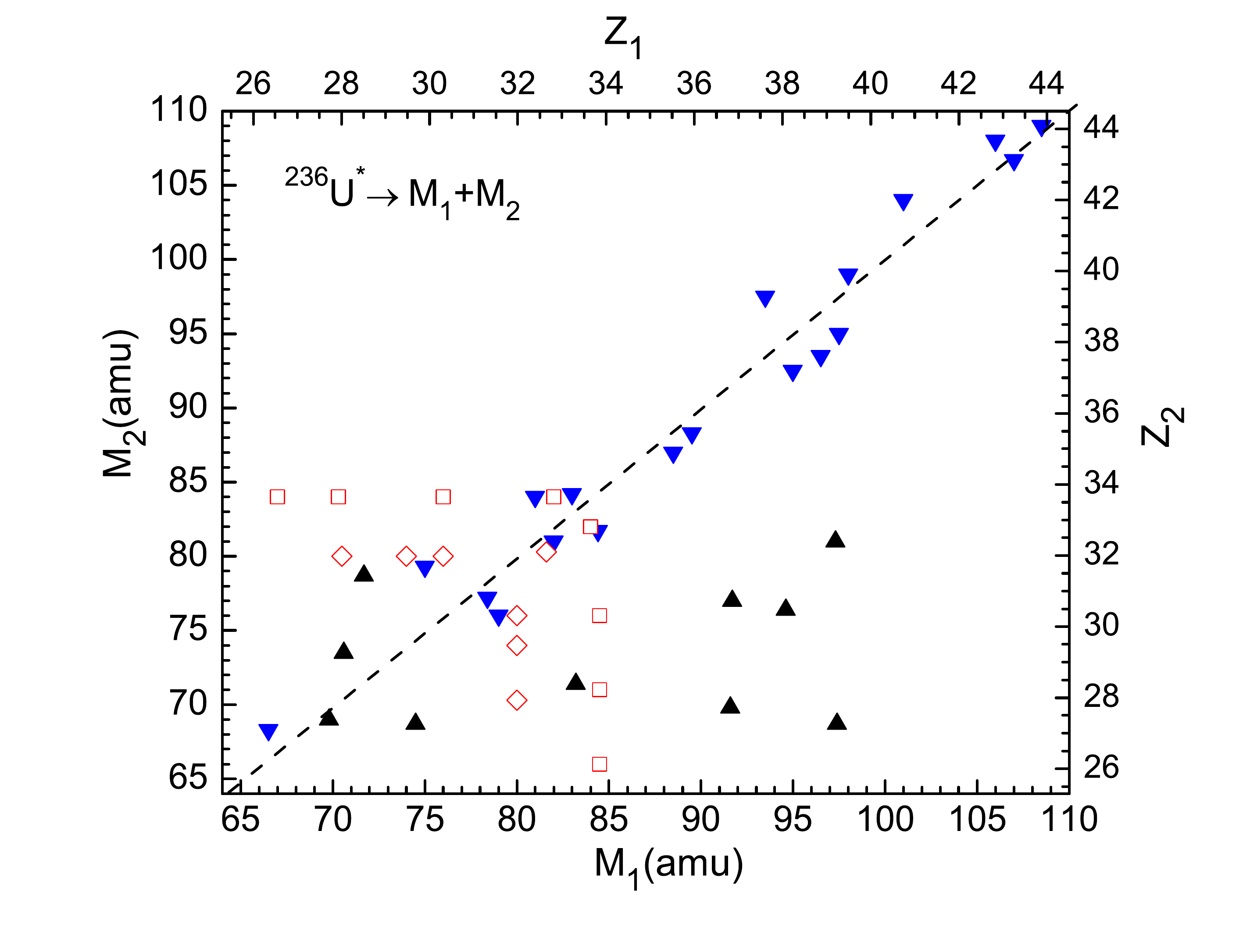}
\vspace*{-1.0 cm}
\caption{\label{CompFFF} Comparison of the maximum values
of the calculated yield of the
collinear cluster tripartition products in the sequential fission
 $^{82}$Ge+($^{154}$Nd $\rightarrow\{^{72}$Ni+$^{82}$Ge,
$^{76}$Zn+$^{78}$Zn\}) (diamonds) and $^{86}$Se+
($^{150}$Ce $\rightarrow\{^{68}$Fe+$^{82}$Ge, $^{72}$Ni+$^{78}$Zn\})
(squares) mechanisms with the experimental data of the mass-mass distribution
of the CCT products in the $^{235}$U($n_{\rm th}$,f)
reaction: with the ones registered in coincidence with approximately equal momenta
(up triangles, the data were taken from Figure 6b of ref.
\cite{PyatkovPHAN73}) and with the ones having approximately equal masses with the
momentum values up to 120 amu(cm ns)$^{-1}$ (down triangles, the data from
Figure 7d of ref.\cite{PyatkovPHAN73}). The charge numbers corresponding
to the presented mass numbers are shown on the top and upper axes.}
\end{figure}

\noindent
\cite{Tashk2011}, which are formed in the primary fission of $^{236}$U (see Fig. \ref{Y236U}).

  The yield of fission products is calculated using the statistical method  based on the driving potentials for
the fissionable system (see Ref. \cite{Tashk2011}). The minima of the potential energy of the decaying system correspond to the charge
numbers of the products, which are produced with large probabilities in the sequential fission of the mother nucleus.

  The probability of the yield of the ternary fission products is relatively large
  in the case of splitting of $^{154}$Nd as the second step of sequential fission.
  The results for the yield of the $^{154}$Nd products are presented in Fig. \ref{Y154Nd}.
  Comparison of the theoretical results for the yield of the true ternary
  fission products (open squares and diamonds) with the observed yields of corresponding mass-mass
  distribution is presented in Fig. \ref{CompFFF}.
 The experimental data (filled up and down triangles)
presented in Fig. \ref{CompFFF} are the mass-mass distribution
 of the $^{235}$U($n_{\rm th}$,f) fission fragments
registered in coincidence by two detectors opposite relative
to the $^{235}$U target.
Different filled symbols correspond to the CCT events, which are selected from the whole data
by different conditions: 1) the CCT products with approximately equal momenta, velocities
(filled up triangles, the results were taken from Fig. 6b of ref.
\cite{PyatkovPHAN73}) and 2) the CCT products with approximately equal masses with the
momentum values up to 120 amu(cm ns)$^{-1}$.
    The results of Ref. \cite{Tashk2011} could interpret only the yield of the   true ternary fission products with comparable masses.
The probability of the yield of the CCT products in the sequential fission channels
of $^{236}$U, which start with the $^{82}$Ge$^*$+$^{154}$Nd$^*$ and   $^{86}$Se$^*$+$^{150}$Ce$^*$
channels are compared in Table 1. From the Table 1 is seen that production of
the Ni isotope is more probable in the fission of $^{154}$Nd.

\begin{figure}
\vspace*{-0.25 cm}
\includegraphics[width=0.80\textwidth]{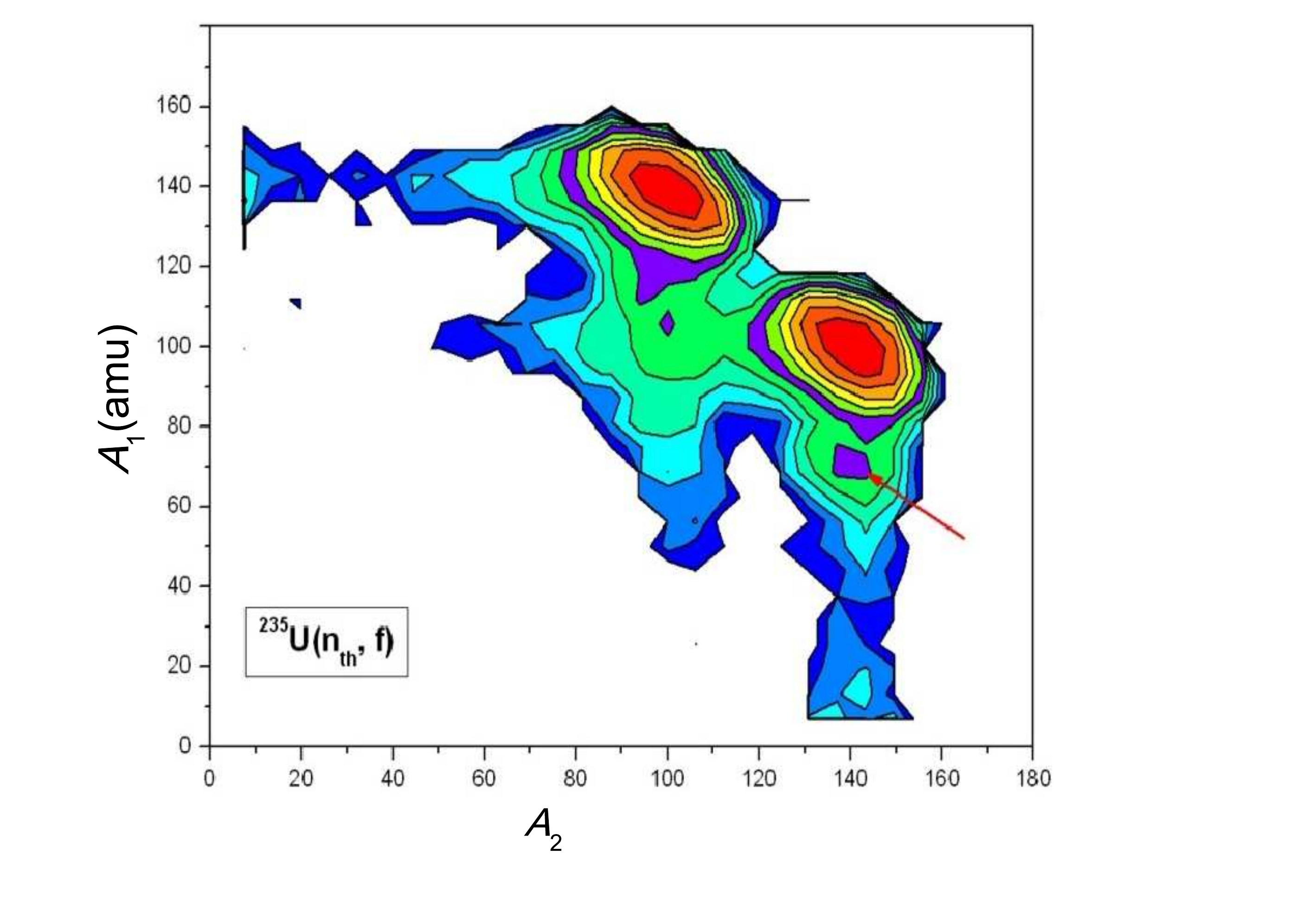}
\vspace*{-1.25 cm}
\caption{\label{Exp236U} Contour map of the mass-mass distribution
(logarithmic scale, with lines approximately a step factor of 1.5) from a coincidence in the two opposite detectors.
The bump in the spectrometer arm1 facing the backing of the U source is
marked by the arrow (from Ref. [6]).}
\end{figure}

But these events do not correspond to the  dominant  processes where the $10^{-3}$
 yields of $^{68-70}$Ni per binary fission  were observed   \cite{Pyatkov10,Pyatkov12}
 with relatively large probability in coincidence with the products with masses $A=$130--150
   in the  $^{235}$U$(n_{\rm th},f)$ (see Fig. \ref{Exp236U}) reaction.

\section{About relation between the two steps of the sequential fission}

The role of the nuclear shell structure is important in the formation
of the ternary system of nuclei and some of them should have
magic or near magic numbers for neutrons or/and protons. This is a
necessary condition for the realization of the true ternary fission
as one channel of the spontaneous fission of $^{252}$Cf. But the existence
of the ternary system as the intermediate state of a system
undergoing to fission is not enough for the occurrence of the true ternary fission.  Three products
can be observed, when  the other massive
super-deformed (residual) part undergoes to fission forming the two other products.
The mechanism of sequential ternary fission with the short time between ruptures of two necks
connecting the
middle cluster 3 to the outer nuclei 1 and 2 may be responsible for the formation of the observed CCT products in Refs. \cite{Pyatkov10,Pyatkov12}. As we stressed above the reason for smallness of the fission probability of the light strongly deformed fragment, which is formed after separation of the tin-like nucleus, is its
large fission barrier ($B_f > $20 MeV). Our calculations showed that
the value of $B_f$ decreases if the heavy tin-like nucleus is not far
from the other residual part of fission:
after the first rupture, its Coulomb field makes a smaller potential well
in the interaction potential between the formed fragments at fission of the
light super-deformed nucleus. In the case of spontaneous fission of
$^{252}$Cf, this super-deformed nucleus is $^{120}$Cd, which can be
considered as a dinuclear system consisting of $^{70}$Ni and $^{50}$Ca.
%\newline
%\begin{center}
%\begin{minipage}{0.80\textwidth}

\begin{figure}
%\vspace*{-0.6 cm}
%\resizebox{0.80\textwidth}{!}{\includegraphics{V1V2NiCaSn}}
\includegraphics[width=0.80\textwidth]{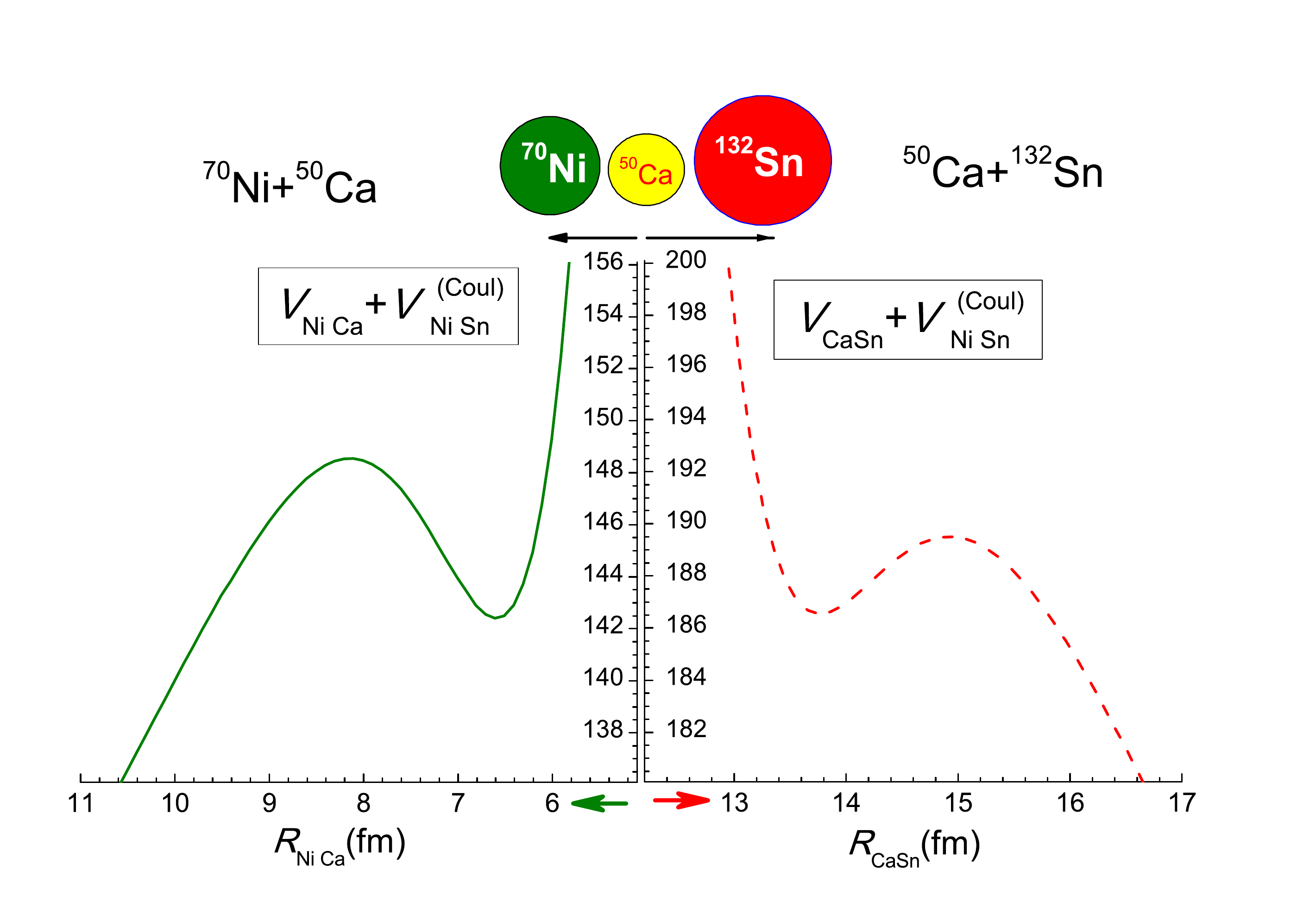}
\vspace*{-0.72 cm}
\caption{\label{V1V2NiCaSn} The nucleus-nucleus interactions between $^{70}$Ni+$^{50}$Ca (left part) and $^{50}$Ca+$^{132}$Sn (right part) of the ternary system  as a function of the relative distances between their mass centres for the collinear configuration.}
\vspace*{-0.7 cm}
\end{figure}
%\end{minipage}
%\end{center}
 The depth of the potential well for the massive system (Ca+Sn)
is smaller than for the one for the light system (Ca+Ni).
In the case of the asymmetric system the decay probability for the heavy fragment is larger since the depth of the potential well is smaller due to larger Coulomb repulsion from the middle cluster:  $Z_1\cdot Z_3 < Z_2\cdot Z_3$ if $Z_1 < Z_2$. It is seen from Fig. \ref{V1V2NiCaSn}, which is calculated  for the configuration $^{70}$Ni+$^{50}$Ca+$^{132}$Sn. Therefore, $^{132}$Sn, or the product close to $^{132}$Sn, is separated as the first product of the fission process. According to the mechanism assumed in our calculations the rupture of the second neck occurs while the first product is  just being accelerated and not far from the dinuclear system consisting of $^{70}$Ni+$^{50}$Ca.
\begin{figure}
\vspace*{-0.5 cm}
\includegraphics[width=0.9\textwidth]{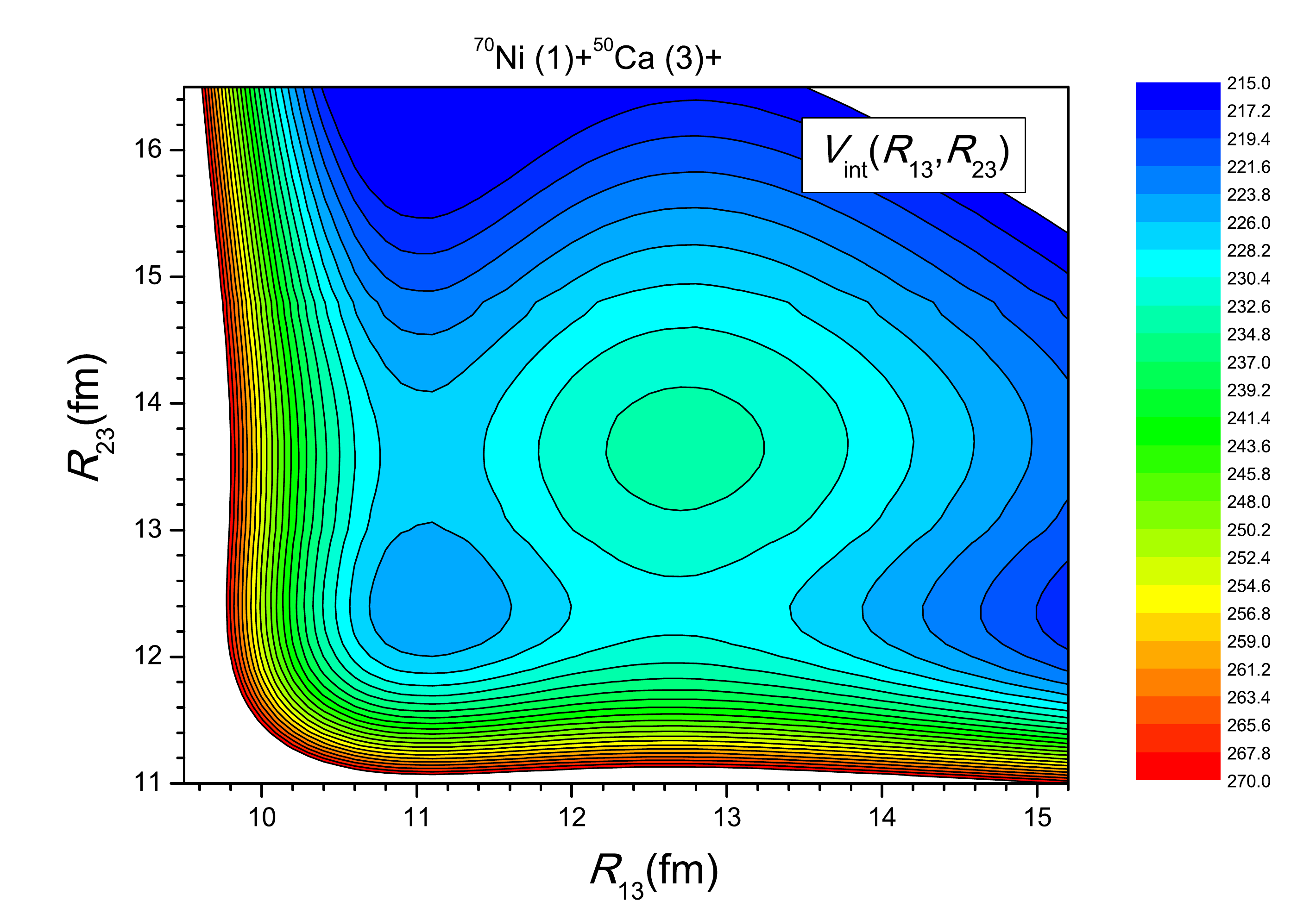}
\vspace*{-0.7 cm}
\caption{\label{NiCaSn3DBarr} The contour plot of the total interaction potential
 $V_{\rm int}$ of the collinear  ternary system $^{70}$Ni+$^{50}$Ca+$^{132}$Sn as function
 of the relative distances $R_{13}$ and $R_{23}$ between the
 centre of the middle cluster and outer nuclei.
$V_{\rm int}=\Sigma_{i=1,2} V_{3i}(R_{3i},Z_i,Z_3,A_i,A_3)+V_{12}^{(\rm Coul)}(Z_1,Z_2,R_{12})$}
\end{figure}
 This is seen from the
contour plot in Fig. \ref{NiCaSn3DBarr}, where the dependence of the  total interaction potential
 $V_{\rm int}$ of the collinear ternary system $^{70}$Ni+$^{50}$Ca+$^{132}$Sn is presented as a function of the relative distances $R_{13}$ and $R_{23}$ between centres of the middle cluster and outer nuclei. This potential includes
 the Coulomb potential $V_{12}^{(\rm Coul)}(Z_1,Z_2,R_{12})$ between the border nuclei ``1'' and ``2'' in addition to the sum of the Coulomb and nuclear interaction between neighbor nuclei (``13'' and  ``32'').
 The minimum of the potential well corresponds to the equilibrium state of the system.
The barrier of the potential  in the direction of the relative distance
``$R_{23}$'' is lower than the barrier in the direction ``$R_{13}$''.
The interaction potentials between
neighbor fragments at the fixed distance between the middle and other fragment are shown
in Fig. \ref{NiCaSn3DBarr}.
Therefore, due to excitation energy
of vibration degrees of freedom the massive fragment, {\it i.e.} Sn, separates first. But the penetration through the barrier $R_{13}$ is possible while
the separated Sn nucleus is not far from the system $^{70}$Ni+$^{50}$Ca.
The realization of this mechanism can explain the observation of the true
ternary fission as the yield of Ni isotopes in coincidence with the massive
product $A\approx 140$ in the experiments of the FOBOS-group \cite{Pyatkov10,Pyatkov12}.
%\begin{center}
   The distances $R_{13}$ and $R_{23}$ between interacting nuclei corresponds to the minimum values of
   the potential wells of
$V_{13}$ and $V_{23}$ interactions, respectively, which are affected by the Coulomb interaction
$V^{(\rm Coul)}_{12}$ of the border fragments (see Fig. \ref{V1V2NiCaSn}).

The procedure of the PES calculation is organized as following. For the given values of charge $Z_3$ and mass $A_3$ numbers of the fragment ``3'' the values of $U$ (PES) are
calculated by varying $Z_1$ from 2 up to 52 and $A_1$ in the wide range
 providing the minimum value of $U$ at the fixed $Z_1$ and $Z_3$ ($A_3$).
The same procedure is made for the range of $A_3$ values  at each
mass and charge configuration of a ternary system.
The interaction potentials $V_{13}$+$V^{(\rm Coul)}_{12}$ and $V_{23}$+$V^{(\rm Coul)}_{12}$ for the ternary system $^{70}$Ni+$^{50}$Ca+$^{132}$Sn  are presented on the left and right part of Fig. \ref{V1V2NiCaSn}. The pre-scission state of the ternary system is determined by the minimum values of the potential wells in
the interaction potentials $V_{ik}(R_{ik})$ in expression for PES Eqn. (\ref{PES3}).

\begin{figure}
\includegraphics[width=0.9\textwidth]{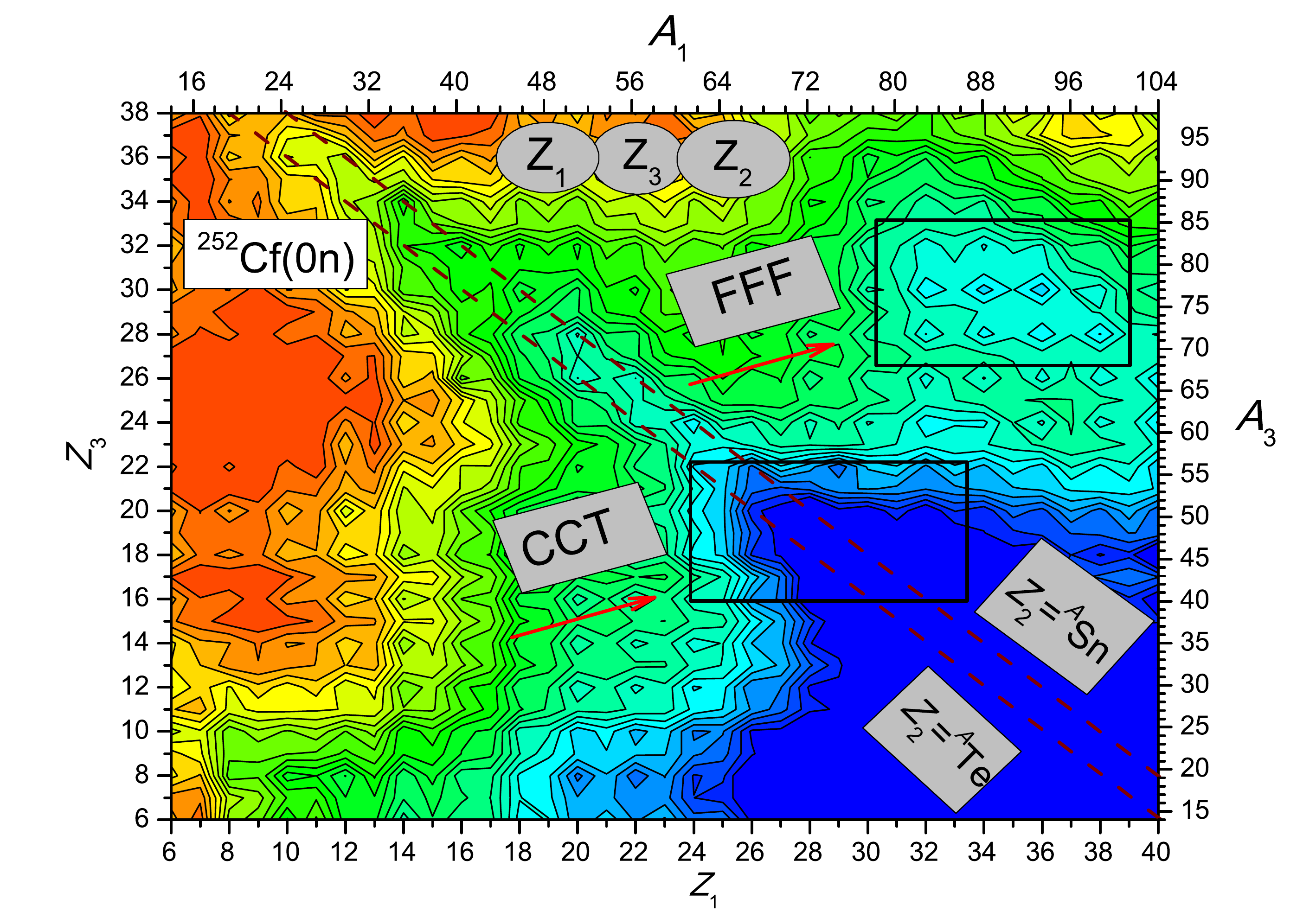}
\vspace*{-0.75 cm}
\caption{\label{PES252Cf} The  contour plots of PES calculated
for the ternary decays characterized by fragments  $Z_1$ ($A_1$)
and $Z_3$ ($A_3$)
 for $^{252}$Cf. The cases with the formation of the isotopes of Tin and Tellurium, $^{A}_{50}$Sn and
 $^{A}_{52}$Te, as the fragments with $Z_2$ are shown by dashed lines. FFF shows  the area of the formation of the three fragments with nearly equal masses.}
\vspace*{-0.25 cm}
\end{figure}

The probability of this configuration is large according to the landscape of PES
calculated for the ternary system  of $^{252}$Cf.
The contour plot of PES for the ternary system formed at the spontaneous fission of $^{252}$Cf is
presented in Fig. \ref{PES252Cf}
 as a function of the charge and mass numbers of fragments ``1'' and ``3''.
The decay is considered with two sequential neck ruptures~\cite{WvOPhysLett14}. The rupture of the necks means overcoming
or tunneling through the barriers, which are illustrated in Fig. \ref{V1V2NiCaSn}.
 It should be noted that the middle cluster is formed as neutron richer in comparison with the border fragments.
 The reason of this theoretical phenomenon is connected with the use of  the
 effective nucleon-nucleon forces suggested by Migdal \cite{Migdal}, which  depend on isospin
 in calculations of the nuclear part of the nucleus-nucleus interaction.
  From the minimizing procedure of the PES by the distribution of neutrons between fragments
   $Z_1, Z_2$, and $Z_3$ we found that the pre-scission configuration $^{70}$Ni+$^{50}$Ca+$^{132}$Sn has lower potential energy in comparison with  those containing the other isotopes ($A \neq 50$) of Ca as the middle cluster \cite{AKNPhysScrip14}.

\begin{figure}
\vspace*{-0.5 cm}
\includegraphics[width=0.9\textwidth]{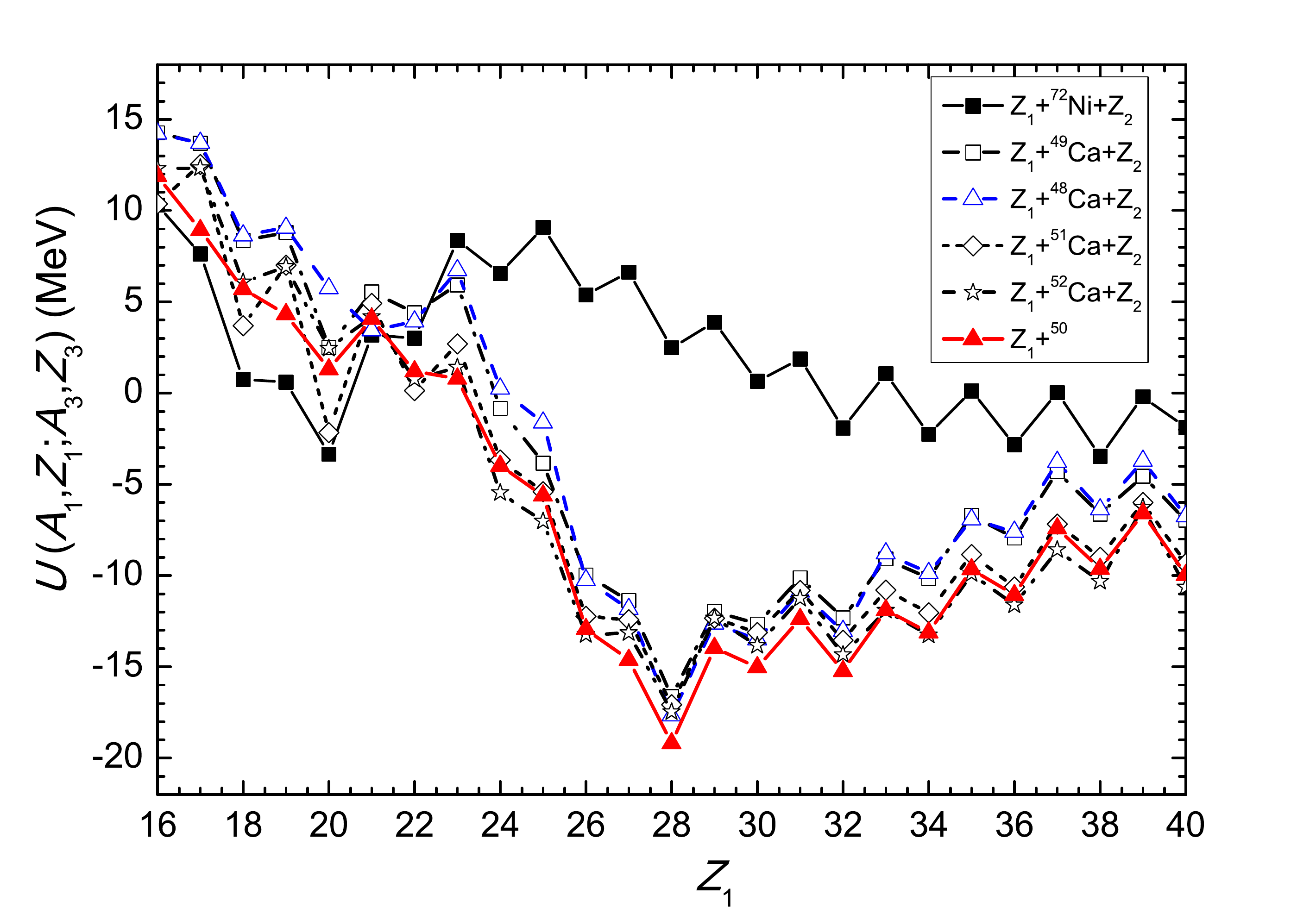}
\vspace*{-0.7 cm}
\caption{\label{UdrNiCaSn}
Comparison of the PES calculated for the pre-scission state of the collinear
ternary system $Z_1+^A$Ca$+Z_2$ formed in the spontaneous fission
of $^{252}$Cf and for the configuration $Z_1+^{72}$Ni$+Z_2$, (black squares) as function of $Z_1$.}
\vspace*{-0.25 cm}
\end{figure}

 This is demonstrated in Fig. \ref{UdrNiCaSn}, where the values of the PES calculated for the  configurations of the ternary system  with different isotopes of Ca (middle cluster) are compared.  In Fig. \ref{UdrNiCaSn} we represent also the curve of results, which were calculated for
 the other case, when
the middle cluster is Ni:  $^{48}$Ca+$^{72}$Ni+$^{132}$S.
The minimum energy of the last configuration is much higher than the one obtained
for the  $^{70}$Ni+$^{50}$Ca+$^{132}$Sn case. This means that the population of the configuration with Ni as the middle cluster is much less probable.

Two important conclusions from the attempt of theoretical analysis of
 of the experimental results \cite{Pyatkov10,Pyatkov12} of the FOBOS-group presented
 in Fig. \ref{Exp236U}  are  1) the correct estimation of the
 ternary system configuration in the pre-scission state and 2) the availability of the
 external Coulomb field causing sequential fission of the super-deformed light residual
 mononucleus accompanying formation of the fragments $A=$130---150 in the first step.

\section{Conclusion}

Results of the PES for the ternary fission of $^{252}$Cf are presented as a binary correlation function of the charge and mass numbers of the middle cluster ($Z_3, A_3$)  and  one of the outer fragments ($Z_1, A_1$).
There are  valleys on the PES corresponding to the formation of
the cluster $Z_2=^{132}$Sn at the different values of $Z_1$ and $Z_3$.
This landscape of PES is related to the long tail in the mass-mass distributions of the experimentally registered products
$M_1$ and $M_2$  demonstrating the persistence of shell structure in the double magic nucleus $^{132}$Sn. The PES contains local minima  showing the favored population of the cluster configurations $^{132}$Sn+$^{50}$Ca+$^{70}$Ni, $^{132}$Sn+$^{38}$S+$^{82}$Ge,
 $^{132}$Sn+$^{36}$Si+$^{84}$Se,  $^{150}$Ba+$^{22}$O+$^{80}$Ge and other
 configurations.
 We found that the middle cluster is more neutron rich than outer fragments.
 The experimentally observed yield of $^{68,70}$Ni isotopes (see Fig. \ref{Exp236U}) 
  is related  to the  configuration $^{132}$Sn+$^{50}$Ca+$^{70}$Ni of the ternary system. The rupture of the neck
 connecting  $^{132}$Sn to the middle cluster takes place earlier  than  the rupture connecting  $^{68,70}$Ni to the $^{50}$Ca.
    The position of the minimum energy on PES for the $^{132}$Sn+$^{72}$Ni+$^{48}$Ca  is much higher
(by 15 MeV)
   than the one for the  configuration $^{132}$Sn+$^{50}$Ca+$^{70}$Ni.
    Therefore,  the small population is observed for the configuration
    with Ni as the middle cluster, and the contribution of this channel
    is observed to be quite small by the FOBOS-group.

\vspace*{-0.77 cm}
\acknowledgments
Authors thank Y. Pyatkov and D. Kamanin for their important discussions of
true ternary fission mechanism and providing us with the data shown in
 the Fig.  \ref{Exp236U}.
A.K.N. is grateful to Profs. Dipak Biswas
for the support and warm hospitality during his staying in India.
W.v.O. thanks the FLNR and BLTP of JINR for their  hospitality extended to him during his stays in Dubna.

\end{document}